\documentclass[oribibl]{llncs}

\usepackage[usenames,dvipsnames]{pstricks}
\usepackage{epsfig}
 \usepackage{pst-grad} 
 \usepackage{pst-plot} 
\usepackage[english]{babel} 
\usepackage[latin2]{inputenc} 
\usepackage{graphicx} 
\usepackage{multirow} 
\usepackage[center]{caption}
\usepackage{geometry}
\usepackage{indentfirst}
\usepackage{savesym}
\usepackage{amsmath}
\savesymbol{iint}
\usepackage{txfonts}
\restoresymbol{TXF}{iint}
\usepackage{ dsfont }
\usepackage{wrapfig}
\usepackage{sidecap}
\usepackage{caption}
\newcommand{\vect}[1]{\mathbf{#1}}    
\newcommand{\mat}[1]{\mathbb{#1}}
\newcommand{\R}{\mathbf{R}}
\newcommand{\Rs}{\mathds{R}}
\newcommand{\g}{\mathbf{H}}
\newcommand{\ig}{\mathbf{H}^{-1}}

\newcommand{\dpart}[2]{\partial_{#1}{#2}}
\newcommand{\pos}{\mathbf{r}}

\newcommand{\X}{\boldsymbol{\chi}}
\newcommand{\Ps}{\boldsymbol{\pi}_{s}}
\newcommand{\Pc}{\boldsymbol{\pi}_{c}}
\newcommand{\Ss}{\boldsymbol{\sigma}_{s}}
\newcommand{\Sc}{\boldsymbol{\sigma}_{c}}
\newcommand{\So}{\boldsymbol{\sigma}_{0}}

\newcommand{\Ep}{\boldsymbol{\epsilon}}
\newcommand{\x}{\mathbf{x}}

\newcommand{\Eun}{\vect{E}_1}

\newcommand{\s}{s}
\newcommand{\Yo}{\mathbf{w}_{0}}
 \newcommand{\cb}{\boldsymbol{\alpha}}         
 \newcommand{\om}{\boldsymbol{\omega}}  
 \newcommand{\vt}{\vect{v}} 
 \newcommand{\m}{\vect{m}} 
 \newcommand{\n}{\vect{p}}  

\newcommand{	\Ade}[2]{\text{Ad}^*_{#1}#2}

\newcommand{	\ade}[2]{\text{ad}^*_{#1}#2}
\newcommand*{\defeq}{\mathrel{\vcenter{\baselineskip0.5ex \lineskiplimit0pt
                     \hbox{\scriptsize.}\hbox{\scriptsize.}}}%
                     =}
 \DeclareMathOperator{\dive}{div}
  \DeclareMathOperator{\curv}{curv}
 \newcommand{\eat}{\ensuremath{\lrcorner \,}}
 
\title{Differential Geometry applied to Acoustics : Non Linear Propagation in
Reissner Beams}
\author{Joel Bensoam}

\institute{Ircam, centre G. Pompidou, CNRS UMR 9912, Acoustic Instrumental Team\\
1 place I. Stravinsky 75004 Paris, France}

\begin{document}
\maketitle

\begin{abstract}

Although acoustics is one of the disciplines of mechanics, its "geometrization" is still limited to a few areas. As shown in the work on nonlinear propagation in Reissner beams, it seems that an interpretation of the theories of acoustics through the concepts of differential geometry can help to address the non-linear phenomena in their intrinsic qualities. This results in a field of research aimed at establishing and solving dynamic models purged of any artificial nonlinearity by taking advantage of symmetry properties underlying the use of Lie groups. The geometric constructions needed for reduction are presented in the context of the "covariant" approach.
\end{abstract}

\section{Introduction}
The Reissner beam is one of the simplest acoustical system that can be treated in the context of mechanics with symmetry. A Lie group is a mathematical construction that handle the symmetry but it is also a manifold on which a motion can take place. As emphasized by Arnold~\cite{Arnold66}, physical motions of symmetric systems governed by the variational principle of least action correspond to geodesic motions on the corresponding group $G$. This paper will try, in a first part, to illustrate this basic concept in the case of the continuous group of motion in space. After a literature survey on this subject, an extension from geodesics to auto-parallel submanifolds is proposed in the second part and naturally leads to the geometric covariant approach available to study evolution problems for fields defined by a variational principle.

\section{Nonlinear model for Reissner Beam}

\subsection{Reissner kinematics }\label{sec:Reissner}
\begin{figure}[h]
\centering
\includegraphics[width=65mm,keepaspectratio=true]{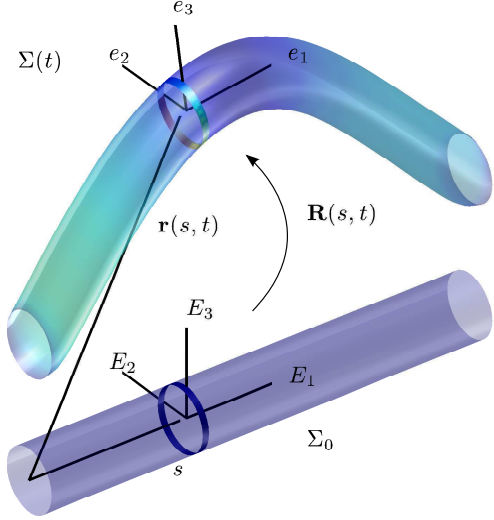}
\caption{Reference and current configuration of a beam. Each section, located at position $\s$ in the reference configuration $\Sigma_0$, is parametrized by a translation $\pos (\s,t)$ and a rotation $\R(\s,t)\in SO_3$ in the current configuration $\Sigma_t$.}
   \label{fig:1}
\end{figure}

A beam of length $L$, with cross-sectional area $A$ and mass per unit volume
$\rho$ is considered.  Following the Reissner kinematics, each section
of the beam is supposed to be a rigid body. The beam configuration can
be described by a position $\pos(\s,t)$ and a rotation $\R(\s,t)$ of
each section. The coordinate $\s$ corresponds to the position of the
section in the reference configuration $\Sigma_0$ (see
figure~\ref{fig:1}).
\subsection{Lie group configuration space}\label{sec:lie}
Any material point $M$ of the beam which is located at
$\x(\s,0)=\pos(\s,0)+\Yo=\s\Eun+\Yo$ in the reference configuration
($t=0$) have a new position (at time $t$) $\x(\s,t)=\pos(\s,t)+\R(\s,t)\Yo$. In other words, the current configuration of the beam $\Sigma_t$ is completely described by a map 
\begin{equation}\label{eq:H}
\begin{pmatrix}
	\x(\s,t)\\
	1
\end{pmatrix}
=\underbrace{
\begin{pmatrix}
	\R(\s,t)&\pos(\s,t)\\
	0&1
\end{pmatrix}
}_{\g(s,t)}
\begin{pmatrix}
	\Yo \\
	1
\end{pmatrix},\quad \R\in SO(3),\quad \pos\in \R^3,
\end{equation}
where the matrix $\g(\s,t)$ is an element of the Lie group $SE(3)=SO(3)\times \R^3$, where $SO(3)$ is the group of all $3\times3$ orthogonal matrices with determinant 1 (rotation in $R^3$). As a consequence, to any motion of the beam a function $\g(s,t)$ of the (scalar) independent variables $s$ and $t$ can be associated. Given some boundary conditions, among all such motions, only a few correspond to physical ones. What are the physical constraints that such motions are subjected to?

In order to formulate those constraints the definition of the Lie algebra is helpful. To every Lie group $G$, we can associate a Lie algebra $\mathfrak{g}$, whose underlying vector space is the tangent space of G at the identity element, which completely captures the local structure of the group. Concretely, the tangent vectors, $\dpart{\s}\g$ and $\dpart{t}\g$,  to the group $SE(3)$ at the point $\g$, are lifted to the tangent space at the identity $e$ of the group. The definition in general is somewhat technical\footnote
{
In the literature, one can find the expression $dL_{g^{-1}}\left(\dot{g}\right)$ where $dL$ stands for the differential of the left translation $L$ by an element of $G$ defined by
\begin{eqnarray*}
L_{g} :G&\rightarrow &G\\
   h&\rightarrow& h \circ g.
\end{eqnarray*}
}
, but in the case of matrix groups this process is simply a multiplication by the inverse matrix $\g^{-1}$. This operation gives rise to definition of two left invariant vector fields in $\mathfrak{g}=\mathfrak{se}(3)$
\begin{eqnarray}\label{eq:Ep}
\hat{\Ep_{c}}(\s,t)&=&\g^{-1}(\s,t)\dpart{\s}{\g}(\s,t)\\
\label{eq:X}
\hat{\X_{c}}(\s,t)&=&\g^{-1}(\s,t)\dpart{t}{\g}(\s,t),
\end{eqnarray}
which describe the deformations and the velocities of the beam. Assuming a linear stress-strain relation, those definitions allow to define a reduced Lagrangian function by the difference of kinetic and potential energy $l(\X_{c},\Ep_{c})=E_{c}-E_{p}$, with
\begin{eqnarray}\label{eq:energy}
 E_{c}(\X_{c})&=&\int_{0}^{L}\frac{1}{2} \X_{c}^T\mat{J}\X_{c} d\s,\\\label{eq:deformation}
  E_{p}(\Ep_{c})&=&\int_{0}^{L}\frac{1}{2} (\Ep_{c}-\Ep_{0})^T\mat{C}(\Ep_{c}-\Ep_{0}) d\s,
\end{eqnarray}
where $\mat{J}$ and $\mat{C}$ are matrix of inertia and Hooke tensor respectively and $\hat{\Ep}_{0}=\g^{-1}(\s,0)\dpart{\s}{\g}(\s,0)$ correspond to the deformation of the initial configuration.
\subsection{Equations of motion}
Applying the Hamilton principle to the left invariant Lagrangian $l$ leads to the 
Euler-Poincar\'e equation
\begin{equation}\label{eq:EP}
\dpart{t}{\Pc}-ad^{*}_{\X_{c}}\Pc=\dpart{\s}{(\Sc-\So)}-ad^{*}_{\Ep_{c}}(\Sc-\So),
\end{equation}
where $\Pc=\mat{J}\X_{c}$ and $\Sc=\mat{C}\Ep_{c}$, (see for example~\cite{roze10},~\cite{Bensoam07a} or ~\cite{Gay-Balmaz09} for details).
In order to obtain a well-posed problem, the compatibility condition, obtained by differentiating ~(\ref{eq:Ep}) and~(\ref{eq:X})
\begin{equation}\label{eq:comp}
\dpart{\s}{\X_{c}}-\dpart{t}{\Ep_{c}}=ad_{\X_{c}}\Ep_{c},
\end{equation}
must be added to the equation of motion. It should be noted that the operators $ad$ and $ad^{*}$ in eq.~(\ref{eq:EP})
\begin{eqnarray}
ad^{*}_{(\om,\vt)}(\m,\n)&=&(\m\times\om+\n\times\vt,\n\times\om)\\
ad_{(\om_{1},\vt_{1})}(\om_{2},\vt_{2})&=&(\om_{1}\times\om_{2},\om_{1}\times\vt_{2}-\om_{2}\times\vt_{1}),
\end{eqnarray}
depend only on the group $SE(3)$ and not on the choice of the particular "metric" $L$ that has been chosen to described the physical problem~\cite{Holm08}. 

Equations~(\ref{eq:EP}) and~(\ref{eq:comp}) are written in material (or left invariant) form ($c$ subscript). Spatial (or right invariant ) form exist also. In this case, spatial variables ($s$ subscript) are introduced by
\begin{eqnarray}\label{eq:Eps}
\hat{\Ep_{s}}(\s,t)&=&\dpart{\s}{\g}(\s,t)\g^{-1}(\s,t)\\
\label{eq:Xs}
\hat{\X_{s}}(\s,t)&=&\dpart{t}{\g}(\s,t)\g^{-1}(\s,t)
\end{eqnarray}
and~(\ref{eq:EP}) leads to the 
conservation law~\cite{Maddocks94}
\begin{equation}\label{eq:momenta}
\partial_{t}\Ps=\partial_{\s}(\Ss-\So)
\end{equation}
where $\Ps=\Ade{\ig}\Pc$ and $\Ss=\Ade{\ig}\Sc$. The $\Ade{}$ map for $SE(3)$ is
\begin{equation}
\Ade{\ig}(\m,\n)=(\R\m+\pos \times \R\n,\R\n).
\end{equation}
Compatibility condition~(\ref{eq:comp}) becomes
\begin{equation}\label{eq:comps}
\partial_{\s}\X_{s}-\partial_{t}\Ep_{s}=ad_{\Ep_{s}}\X_{s}.
\end{equation}
Equations~(\ref{eq:EP}) and~(\ref{eq:comp}) (or alternatively (~\ref{eq:momenta}) and~(\ref{eq:comps})) provide the exact non linear Reissner beam model and can be used to handle the behavior of the beam if the large displacements are taking into account.

Notations and assumptions vary so much in the literature, it is often difficult to recognize this model (see for example~\cite{Celledoni} for a formulation using quaternions). However, this generic statement is used to classify publications according to three axes. In the first one, the geometrically exact beam model is the basis for numerical formulations. Starting with the work of Simo~\cite{simo85}, special attention is focused on energy and momentum conserving algorithms~\cite{Simo95},~\cite{Leyendecker}. Numerical solutions for planar motion are also investigated in~\cite{Gams07}. Even, in some special sub-cases (namely where the longitudinal variables do not appear) the non-linear beam model gives rise to linear equations which can be solved by analytical methods~\cite{Bishop04}. 

Secondly, much of the literature is also devoted to the so-called Kirchhoff's rod model. In this case, shear strain is not taken into account along a thin rod (i.e., its cross-section radius is much smaller than its length and its curvature at all points). In this approximation cross-sections are perpendicular to the central axis of the filament and the rotation matrix can be given in the Frenet-Serret frame. (see ~\cite{Fonseca03} , ~\cite{Goriely97}, ~\cite{Argeri09}, for example). In that context an interesting geometric correspondence between Kirchhoff rod and Lagrange top can be made~\cite{Nizette99}.


Finally, if only rigid motion is investigated, ( i.e. if the spatial dependence in~(\ref{eq:EP}) is canceled: $\dpart{\s}{}\equiv 0$) the so-called underwater vehicle\footnote{underwater vehicle in the case that the center of buoyancy and the center of gravity are coincident} model is obtained. In absence of exterior force and torque, the equation of motion for a rigid body in an ideal fluid become more simply ~\cite{Leonard97},~\cite{Holmes98}
\begin{equation}\label{eq:UV}
\dpart{t}{\Pc}=ad^{*}_{\X_{c}}\Pc, \text{ that is}
\begin{cases}
	\dot{\m}=\m\times\om+\n\times\vt\\
	\dot{\n}=n\times\om
\end{cases}
\end{equation}
In this simpler form, a geometric interpretation is easier. The solution of the equation of motion mentioned above, if it exists,  should be interpreted as a geodesic of the group $SE(3)$ endowed with a non-canonical left invariant metric $\mat{J}$. To accomplish the correspondence between the Euler-Poincar\'e's equation and geodesic equation the historical definition of the covariant derivative is exposed in the next section.

\section{Geometric interpretation}

\subsection{Geodesics on curved spaces}
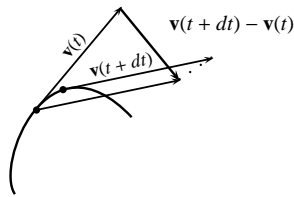
\begin{SCfigure}
  \centering
\scalebox{.8} 
{
\begin{pspicture}(0,-1.6973269)(6.46,1.7254488)
\usefont{T1}{ptm}{m}{n}
\rput{12.335343}(0.14328112,-0.44054615){\rput(2.11,0.46267313){\footnotesize $\mathbf{v}(t+dt)$}}
\psbezier[linewidth=0.04](0.34,-1.6773269)(0.0,-1.1973269)(0.88,1.0826731)(2.28,-0.3973269)
\psline[linewidth=0.03cm,arrowsize=0.05291667cm 2.0,arrowlength=1.4,arrowinset=0.4]{->}(0.68,-0.27732688)(2.12,1.4426731)
\psline[linewidth=0.03cm,arrowsize=0.05291667cm 2.0,arrowlength=1.4,arrowinset=0.4]{->}(1.24,0.082673125)(3.64,0.58267313)
\usefont{T1}{ptm}{m}{n}
\rput{51.141293}(1.1087453,-0.7518296){\rput(1.34,0.8026731){\footnotesize $\mathbf{v}(t)$}}
\usefont{T1}{ptm}{m}{n}
\rput(3.94,1.1476731){$\mathbf{v}(t+dt)-\mathbf{v}(t)$}
\psline[linewidth=0.04cm,arrowsize=0.05291667cm 2.0,arrowlength=1.4,arrowinset=0.4]{<-}(3.08,0.20267312)(2.08,1.4226731)
\psline[linewidth=0.04cm,linestyle=dotted,dotsep=0.16cm](3.06,0.22267312)(3.62,0.6026731)
\psdots[dotsize=0.12](0.7,-0.27732688)
\psdots[dotsize=0.12](1.14,0.06267312)
\psline[linewidth=0.03cm,arrowsize=0.05291667cm 2.0,arrowlength=1.4,arrowinset=0.4]{->}(0.7,-0.27732688)(3.1,0.22267312)
\end{pspicture} 
}
\caption{For flat manifolds, a trivial parallel transport is used to compute the acceleration.}
   \label{fig:2}
\end{SCfigure}

\begin{SCfigure}
       \captionsetup{labelsep=space,justification=justified,singlelinecheck=off}
	\caption{Parallel transport along a curve: Let $\cb(t)$ be a curve on a surface $S$ and consider the envelope of the family of tangent planes of $S$ along $\cb$ (see figure~\ref{fig:3}). Assuming that $\cb(t)$ is nowhere tangent to an asymptotic direction, this envelope is a regular surface $\Sigma$ which is tangent to $S$ along $\cb$. Thus, the parallel transport along $\cb$ of any vector $\vect{w}\in T_{p}(S)$, $p\in S$, is the same whether we consider it relative to $S$ or to $\Sigma$.
Furthermore,  $\Sigma$ is a developable surface; hence can be mapped by an isometry $\phi$ into a plane $P$ (without stretching or tearing). Parallel transport of a vector $\vect{w}$ is then obtained using usual parallel transport in the plane along $\phi(\cb)$ and pull it back to $\Sigma$ (by $d\varphi^{-1}$).}
  \centering
  \includegraphics[width=0.5\textwidth]{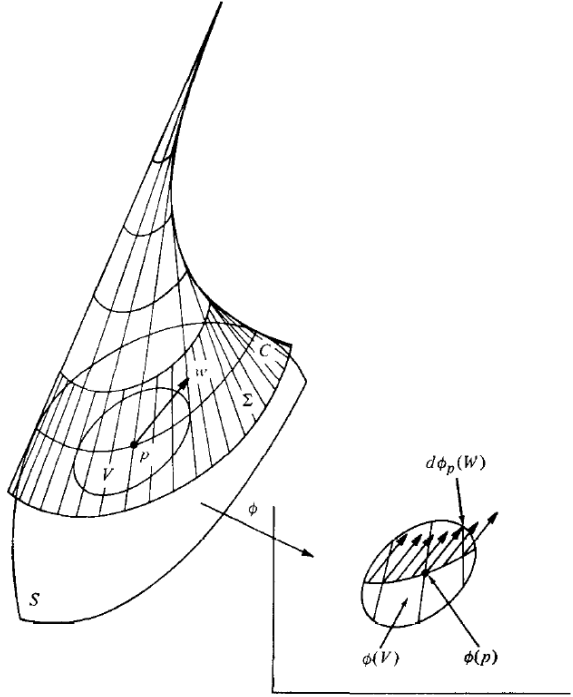}
   \label{fig:3}
\end{SCfigure}

A trajectory of a particle of mass m which is moving on a manifold\footnote{a surface for short} $M$ can be thought as a curve $\cb(t)$ on $M$ and $\vect{v}(t)=\dot{\cb}(t)$ is the speed of the particle. According to the Newton's second Law of motion, its acceleration (the variation of its velocity) is proportional to the net force acting upon it $\sum \vect{F}=m\frac{d\vect{v}}{dt}$. The expression of this variation, $\vect{v}(t+dt)-\vect{v}(t)$, shows that the velocities are evaluated at two different points of the curve: $\cb(t+dt)$ and $\cb(t)$ which are, \textit{a priori}, incommensurable quantities. So, one of the two vectors needs to be parallel transport as it is  illustrated, for flat manifolds, in figure~(\ref{fig:2}). 
For curved manifolds the operation is not so easy and its historical construction is related by M.P. do Carmo in~\cite{doCarmo76} for surfaces of $R^3$ (see figure~\ref{fig:3}).  
Technically, this historical construction gives rise to the concept of the covariant derivative $\frac{D\vect{w}}{dt}=\nabla_{\vect{v}}\vect{w}$ of a vector field $\vect{w}$ along $\cb$. The parametrized curves $\cb : I \rightarrow R^2$ of a plane along which the field of their tangent vector $\vect{v}(t)$ is parallel are precisely the straight lines of that plane. The curves that satisfy an analogous condition , i.e. 
\begin{equation}\label{eq:geodesic}
\frac{D\vect{v}}{dt}=\nabla_{\vect{v}}\vect{v}=0,
\end{equation} 
for a surface are called geodesics. Intuitively, the acceleration as seen from the surface  vanishes :  in absence of net force, the particle goes neither left nor right, but straight ahead. 

The kinetic energy~(\ref{eq:energy}) define a left invariant Riemannian metric on $SE(3)$, and 
then define also a symmetric connection $\nabla$ which is compatible with this metric (Levi-Civita connection). It can be shown that geodesic equation~(\ref{eq:geodesic}) for this particular connection coincide with Euler-Poincar\'e equation of motion~(\ref{eq:UV}) when $SE(3)$ is endowed with the kinetic metric~(\ref{eq:energy}).

Now, this equation deals with motion of rigid body described by a single scalar variable $t$. So what is the geometric interpretation of the equations of motion~(\ref{eq:EP}) and~(\ref{eq:comp}) for which two variables $s$ and $t$ are involved. In other word, can we extend a geodesic, which is a 1-dimensional manifold, to 2-dimensional geodesic ?

\subsection{Auto-Parallel submanifolds, covariant point of view}
A geodesic curve on a surface $S$ is a 1-dimensional submanifold of $S$ for which the parallel transport of its initial velocity stay in its own tangent space. In that sense, a geodesic is an auto-parallel curve. If now, geodesics are seen as auto-parallel curves on surface, a definition of an n-dimensional auto-parallel submanifolds can be made. 

A submanifold $M$ is auto-parallel in $S$ if  the parallel translation of any  tangent vector of $M$ along any curve in $M$  stays in its own tangent space $T(M)$. Note that a parallel translation of a vector $\vect{w}\in T(M)$ certainly belongs  to $T(S)$ but not necessarily to $T(M)$. In other words, $M$ is auto-parallel in $S$ with respect to the connection $\nabla$ of $S$ if $\nabla_{\vect{X}}\vect{Y}$ belongs to $T(M)$, $\forall \vect{X},\vect{Y}\in T(M)$. 
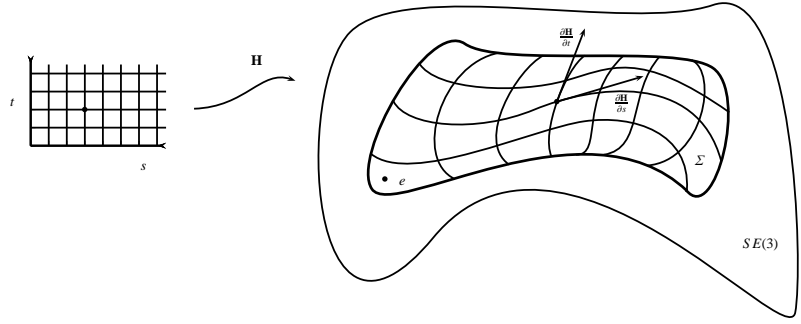
\begin{figure}[h]
\begin{center}
\scalebox{.6} 
{
\begin{pspicture}(0,-4.6239285)(17.888395,4.6239285)
\psbezier[linewidth=0.04](9.345833,2.6960716)(9.725833,2.0560715)(11.585834,2.0160716)(12.585834,2.2960715)(13.585834,2.5760715)(14.205833,3.0160716)(16.225834,1.5960716)
\psbezier[linewidth=0.04](8.785833,1.8398362)(9.065833,1.2734834)(10.607778,1.1205422)(11.905833,1.6560715)(13.20389,2.191601)(15.505834,2.5160716)(16.045834,0.5160716)
\psbezier[linewidth=0.04](8.285833,0.6760716)(9.165833,-0.04392847)(11.885834,1.2360716)(13.005834,1.3960716)(14.125833,1.5560715)(15.365832,1.1360716)(15.285832,-0.12392837)
\psbezier[linewidth=0.04](10.145833,0.11607163)(9.205833,0.5560716)(9.825833,2.6760716)(11.1258335,2.9160717)
\psbezier[linewidth=0.06](15.285832,-0.14392847)(13.045834,1.9760716)(8.785833,-0.8239283)(8.305833,-0.10392837)(7.8258333,0.6160715)(9.562756,3.6639998)(10.365833,3.1160717)(11.168911,2.5681436)(15.082056,3.1310015)(15.865832,2.5760715)(16.64961,2.0211415)(15.912596,-0.9508111)(15.265834,-0.12392837)
\psbezier[linewidth=0.04](15.865832,3.9760718)(17.705833,4.5960717)(17.868395,-1.9046589)(17.685833,-2.8439283)(17.503271,-3.7831976)(12.485833,2.2360716)(9.665833,-1.1839285)(6.8458333,-4.6039286)(6.5928965,2.1482148)(7.8258333,3.3760717)(9.05877,4.6039286)(14.025832,3.356072)(15.865832,3.9760718)
\usefont{T1}{ptm}{m}{n}
\rput(12.615833,3.2610717){$\frac{\partial\mathbf{H}}{\partial t}$}
\psdots[dotsize=0.12](8.592896,0.11607153)
\usefont{T1}{ptm}{m}{n}
\rput(8.985833,0.06107173){$e$}
\psbezier[linewidth=0.04](11.345834,0.43607154)(10.585834,0.9760716)(11.525833,2.5560715)(12.105833,2.8560715)
\psbezier[linewidth=0.04](12.285833,0.6160716)(12.045834,1.1560715)(12.605833,2.5760715)(13.025833,2.8560715)
\psdots[dotsize=0.12](12.405833,1.8360716)
\psline[linewidth=0.04cm,arrowsize=0.05291667cm 2.0,arrowlength=1.4,arrowinset=0.4]{->}(12.425834,1.8360716)(14.325833,2.3960717)
\psline[linewidth=0.04cm,arrowsize=0.05291667cm 2.0,arrowlength=1.4,arrowinset=0.4]{->}(12.405833,1.8360716)(13.025833,3.4560716)
\usefont{T1}{ptm}{m}{n}
\rput(13.835834,1.8010715){$\frac{\partial\mathbf{H}}{\partial \s}$}
\psbezier[linewidth=0.04](14.405833,0.4160716)(15.665833,0.5160716)(15.985832,2.4974275)(15.385834,2.7560716)
\psbezier[linewidth=0.04](12.825833,0.6560716)(13.445833,0.6360715)(12.865833,1.9160715)(13.885834,2.8560715)
\psbezier[linewidth=0.04](13.445833,0.6560716)(14.345834,0.801477)(13.965834,2.027423)(14.685834,2.8360715)
\usefont{T1}{ptm}{m}{n}
\rput(16.915833,-1.3389283){$SE(3)$}
\usefont{T1}{ptm}{m}{n}
\rput(15.575832,0.5410716){$\Sigma$}
\rput(0.74583334,0.85607153){\psaxes[linewidth=0.04,labels=none,ticksize=0.10583333cm,dx=0.4cm,dy=0.4cm]{-<}(0,0)(0,0)(3,2)}
\psline[linewidth=0.04cm](1.1458333,0.85607153)(1.1458333,2.6560717)
\psline[linewidth=0.04cm](1.5458333,0.85607153)(1.5458333,2.6560717)
\psline[linewidth=0.04cm](1.9458332,0.85607153)(1.9458332,2.6560717)
\psline[linewidth=0.04cm](2.3458333,0.85607153)(2.3458333,2.6560717)
\psline[linewidth=0.04cm](2.7458332,0.85607153)(2.7458332,2.6560717)
\psline[linewidth=0.04cm](3.1458333,0.85607153)(3.1458333,2.6560717)
\psline[linewidth=0.04cm](3.545833,0.85607153)(3.545833,2.6560717)
\psline[linewidth=0.04cm](0.74583334,2.4560714)(3.7458336,2.4560714)
\psline[linewidth=0.04cm](0.74583334,2.0560718)(3.7458336,2.0560718)
\psline[linewidth=0.04cm](0.74583334,1.6560718)(3.7458336,1.6560718)
\psline[linewidth=0.04cm](0.74583334,1.2560717)(3.7458336,1.2560717)
\psdots[dotsize=0.12](1.9458332,1.6560718)
\usefont{T1}{ptm}{m}{n}
\rput(3.2358332,0.38107154){$\s$}
\usefont{T1}{ptm}{m}{n}
\rput(0.33583328,1.8210717){$t$}
\psbezier[linewidth=0.04,arrowsize=0.05291667cm 2.0,arrowlength=1.4,arrowinset=0.4]{->}(4.3658333,1.6760714)(5.5258336,1.7125932)(5.8528967,2.5560715)(6.6128964,2.2960715)
\usefont{T1}{ptm}{m}{n}
\rput(5.7558336,2.7610717){$\mathbf{H}$}
\end{pspicture} 
}
\caption{Symbolic representation of a parametrized surface $\Sigma$ immersed into the group $G=SE(3)$}
   \label{fig:5}
   \end{center}
\end{figure}
A correspondence between auto-parallel surface and solutions to equations~(\ref{eq:EP}) and~(\ref{eq:comp}) is still to be demonstrated. In this case, any motion of the beam must be seen as a map from $[0,L]\times \R\subset \Rs^2$ to $SE(3)$ given by
\begin{equation}
\label{eq:cov}
   (\s,t)\rightarrow \g(s,t),
\end{equation}
rather then a curve $t \rightarrow \g(.,t)$ in the infinite dimensional configuration space $\mathcal{F}([0; L]; SE(3))$ of functions from $[0; L]$ to $SE(3)$. In this perspective, solving a physical variational problem is therefore transposed to the problem of finding an auto-parallel immersed surface as it is illustrated symbolically in figure~(\ref{fig:5}). This process illustrates the covariant (as opposed to dynamical) formulation of a variational problem (see~\cite{ellis2011lagrange},~\cite{Lopez:2008}).

More precisely, the map~(\ref{eq:cov}) should be interpreted as a (local) section $\mathfrak{s}(\vect{x})=\left(\vect{x},\g(\vect{x})\right)$ of the principal fiber bundle $P\rightarrow X$ with structure group $G=SE(3)$
\begin{equation*}
\pi : P=X\times SE(3)\rightarrow X ,\quad \pi (\vect{x},\g ) \defeq \vect{x}
\end{equation*}
over the spacetime $X = [0; L]\times \Rs$, $(s, t) = \vect{x}$. The Lagrangian is then defined in the phase bundle $L : J^1P\rightarrow \Rs$, 
where $J^1P$ denotes the first jet bundle of the bundle $P$. If  $L$ is invariant under the action of $G$, the variational principle drops to the quotient space $(J^1P)/G$. This quotient is an affine bundle on $X$ which can be identified to the bundle of connections $C\rightarrow X$. It induces a reduced lagrangian $l : C\rightarrow X$ from $L$ and a reduced section $\bar{\mathfrak{s}}\in\Gamma(C)$ from $\mathfrak{s}\in\Gamma(P)$. In that context, the multidimensional generalization of the equations of motion~(\ref{eq:EP}), compatibility~(\ref{eq:comp}) and conservation law~(\ref{eq:momenta}) are formulated by M. Castrill\'on~L\'opez in~\cite{Lopez:2001}. The equation of motion~(\ref{eq:EP}) now yields
\begin{equation} \label{eq:G}
\dive{\frac{\delta l}{\delta \bar{\mathfrak{s}}}}+\ade{\bar{\mathfrak{s}}}{\frac{\delta l}{\delta \bar{\mathfrak{s}}}}=0
\end{equation}
where $\dive$ stands for the divergence operator defined by the volume form $v$ (here, $v=ds\wedge dt$). The compatibility condition~(\ref{eq:comp})gives rise to the flatness of $\bar{\mathfrak{s}}$ (integrability condition)
\begin{equation}\label{eq:Gcomp}
\curv(\bar{\mathfrak{s}})=d\bar{\mathfrak{s}}+[\bar{\mathfrak{s}},\bar{\mathfrak{s}}]=0.
\end{equation}
Finally, introducing the Cartan-Poincar\'e\footnote{$n$ is the dimension of the base manifold $X$, here $n=2$.} n-form, $\Theta_{L}$, the symmetries of a variational problem produces conservation laws by means of the Noether's Theorem
\begin{equation}\label{eq:Gmomenta}
d(J^1\mathfrak{s})^*\mathcal{J}=0.
\end{equation}
The form $\mathcal{J}$ induces a conserved quantity  since its differential vanishes along the critical section $\mathfrak{s}$. It can be understood as a current form (like in electromagnetism). In that sense, this formulation is more appropriated to describe a conservation law then the partial derivative balance law~(\ref{eq:momenta}). But for a non-specialist audience, the definition of this form
\begin{equation*}
\mathcal{J}(\xi)=((\xi)^*)^{(1)}\eat\Theta_{L},\quad \forall \xi\in \mathfrak{g}
\end{equation*}
is quite obscure (in particular the relationship between the Cartan-Poincar\'e form and the conserved quantity). Here $\eat$ stands for the interior product and $\xi^*$ is related to the infinitesimal vector field generated by the symmetry\footnote{
More precisely, 
Given an element $\xi$ of the Lie algebra $ \mathfrak{g}$ of $G$, the infinitesimal generator of the
G-action on $P$ is denoted by $\xi^*\in \mathfrak{X}(P)$, that is $\xi^*_p=d(p \exp t\xi)/dt |_{t=0}$ for any $p\in P$
}.

\section{Conclusion}
 A geometrical approach of the dynamic of a Reissner beam has been studied in this article in order to take into account non linear effects due to large displacements. There are basically two different geometric approaches available to study evolution problems for fields defined by a variational principle. The first approach, called the "dynamical" approach, uses, as its main ingredient, the infinite dimensional manifold as configuration space (TQ). The reduction techniques developed in the dynamical framework have been studied thoroughly in the literature (see for example~\cite{marsden:1999} and the references therein cited), but it presents the difficulty to handle geodesic curves in an infinite dimensional function space.

As an alternative, the covariant formulation allows to consider a finite dimensional configuration space (the dimension of the symmetry group itself). Although its roots go back to De Donder~\cite{de1930theorie}, Weyl~\cite{weyl1935geodesic}, Caratheodory ~\cite{caratheodory1999calculus}, after J. M. Souriau in the seventies~\cite{Souriau:1970}, the classical field theory has  been only well understood in the late 20th century (see for example~\cite{kanatchikov1998canonical} for an extension from symplectic to multisymplectic form). It is therefore not surprising that, in this covariant or jet formulation setting, the geometric
constructions needed for reduction have been presented even more recently. In that circumstances in the literature, it is also not easy to understand how the multisymplectic form can be obtained from the differential of the Cartan-Poincar\'e n-form, which is crucial to give rise to an Hamiltonian framework (Lie-Poisson Schouten-Nijenhuis (SN) brackets~\cite{Schouten1940}). An understandable theory, that can unify all the results obtained "ad hoc", case by case, is still missing to our knowledge.


\end{document}